# Intelligent Advisory System for Supporting University Managers in Law

A. E. E. ElAlfi
Dept. of Computer Science
Mansoura University
Mansoura Egypt, 35516
Ael_Alfi@yahoo.com

M. E. ElAlami
Dept. of Computer Science
Mansoura University
Mansoura Egypt, 35516
Moh_ElAlmi@mans.eun.eg

*Abstract*— The rights and duties of both staff members and students are regulated by a large and different numbers of legal regulations and rules. This large number of rules and regulations makes the decision-making process time consuming and error boring. Smart advisory systems could provide rapid and accurate advices to managers and give the arguments for these advices. This paper presents an intelligent advisory system in law to assist the legal educational processes in universities and institutes. The aims of the system are:
to provide smart legal advisors in the universities and institutes, to integrate rules and regulations of universities and institutes in the e-government, to ease the burden on the legal advisor and the provision of consulting services to users, to achieve accurate and timely presentation of the legal opinion to a given problem and to assure flexibility for accepting changes in the rules and legal regulations. The system is based on experienced jurists and the rules and regulations of the law organizing Saudi Arabia universities and institutes.

## I. INTRODUCTION

Decision making, often viewed as a form of reasoning towards action, has raised the interest of many scholars including philosophers, economists, psychologists, and computer scientists for a long time. Any decision problem aims to select the "best" or sufficiently "good" action(s) that are feasible among different alternatives, given some available information about the current state of the world and the consequences of potential actions [1]. Advisory systems provide the advices and assist for solving problems that are normally solved by human experts. They can be classified as a type of expert systems [2,3]. Both advisory systems and expert systems are problem-solving packages that mimic a human expert in a special area. These systems are constructed by eliciting knowledge from human experts and coding it into a form that can be used by a computer in the evaluation of alternative solutions to problems within that domain of expertise. Advisory systems do not make decisions but rather help guide the decision maker in the decision-making process, while leaving the final decision-making authority up to the human user [4]. The decision maker works in collaboration with the advisory system to identify problems that need to be addressed, and to iteratively evaluate the possible solutions to unstructured decisions. For example, a manager of a firm could use an advisory system that helps assess the impact of a management decision on firm value [5] or an oncologist can use an advisory system to help locate brain tumors [6]. In these two examples, the manager and the oncologist are ultimately (and legally) accountable for any decisions/diagnoses made. Traditionally rule-based expert systems operate best in structured decision environments, since solutions to structured problems have a definable right answer, and the users can confirm the correctness of the decision by evaluating the justification provided by explanation facility [7]. Luger [8] has presented some limitations of current expert systems.

Advisory systems are designed to support decision making in more unstructured situations which have no single correct answer. In unstructured situations cooperative advisory systems that provide reasonable answers to a wide range of problems are more valuable and desirable than expert systems that produce correct answers to a very limited number of questions [9].

Advisory systems support decisions that can be classified as either intelligent or unstructured, and are characterized by novelty, complexity, and open-endedness [10]. In addition to these characteristics, contextual uncertainty is ubiquitous in unstructured decisions, which when combined exponentially increases the complexity of the decision-making process. Because of the novel antecedents and lack of definable solution, unstructured decisions require the use of knowledge and cognitive reasoning to evaluate alternative courses of action to find one that has the highest probability of desirable outcome [11]. The more context-specific knowledge acquired by the decision maker in these unstructured decision-making situations, the higher the probability that they will achieve the desirable outcome [4].

The decision-making process that occurs when users utilize advisory systems is similar to that which is used for the judge-advisor model developed in the organizational behavior [12,13]. Under this model, there is a principle decision maker that solicits advice from many sources. However, the decision maker “holds the ultimate authority for the final decision and is made accountable for it” [14]. The judge-advisor model suggests that decision makers are motivated to seek advice from others for decisions that are important, unstructured, and involve uncertainty.

Universities made great strides in many areas related to e-government systems, but legal advice to decision makers in universities is still depending largely on the legal advisors .Fortunately, the law rules are considered as fertile ground for building knowledge based systems that can serve as high-level advisory in law [15].

The paper is organized as follows:

Section 2 presents the system design and development. Section 3 presents case study. Section 4 is devoted to system flexibility and merits. The paper is terminated by concluding remarks and perspectives summarizing the obtained results and proposing problems for future work.

## II. System Design and Development

The intelligent advisory system (IAS) must provide assistance for the decision making process. Its aim is to capture the expertise in a form that others can use, and to act as an operational guide without limiting the independent exploration of the user.

The three main processes in advisory systems are knowledge acquisition, cognition, and interface. The user interface allows users to access the IAS and includes multiple windows to visualize how the main parameters interrelate with each other. Input data such as certificates, student grade, age etc., are introduced through the user interface. After input details have been entered, detailed output parameters such as, student accepted or rejected, are displayed. Advice messages are provided to the decision maker during the decision making process. They indicate the next action to be performed every time the IAS program is executed. These messages appear on windows until the decision making process constraints are satisfied.

### *A. Proposed system architecture and design*

The iterative support of advisory systems in the decision-making process is shown in figure1. Knowledge is acquired by knowledge engineers from the experts and the documents of rule and regulations. The cognition is inferred by inference engine. The system has a monitoring agent to identify the need for identifying unstructured decisions that need to be addressed. Decision maker uses the user interface to communicate with the system. There exist an explanation facility to display the arguments of any decision. These are displayed in figure 1 as the flow of information from domain variables to the inference engine. If environmental domain variables exceed expected norms, then the system will notify the user that there is a situation which needs to be addressed and will begin the iterative decision-making process by offering a suggested course of action.

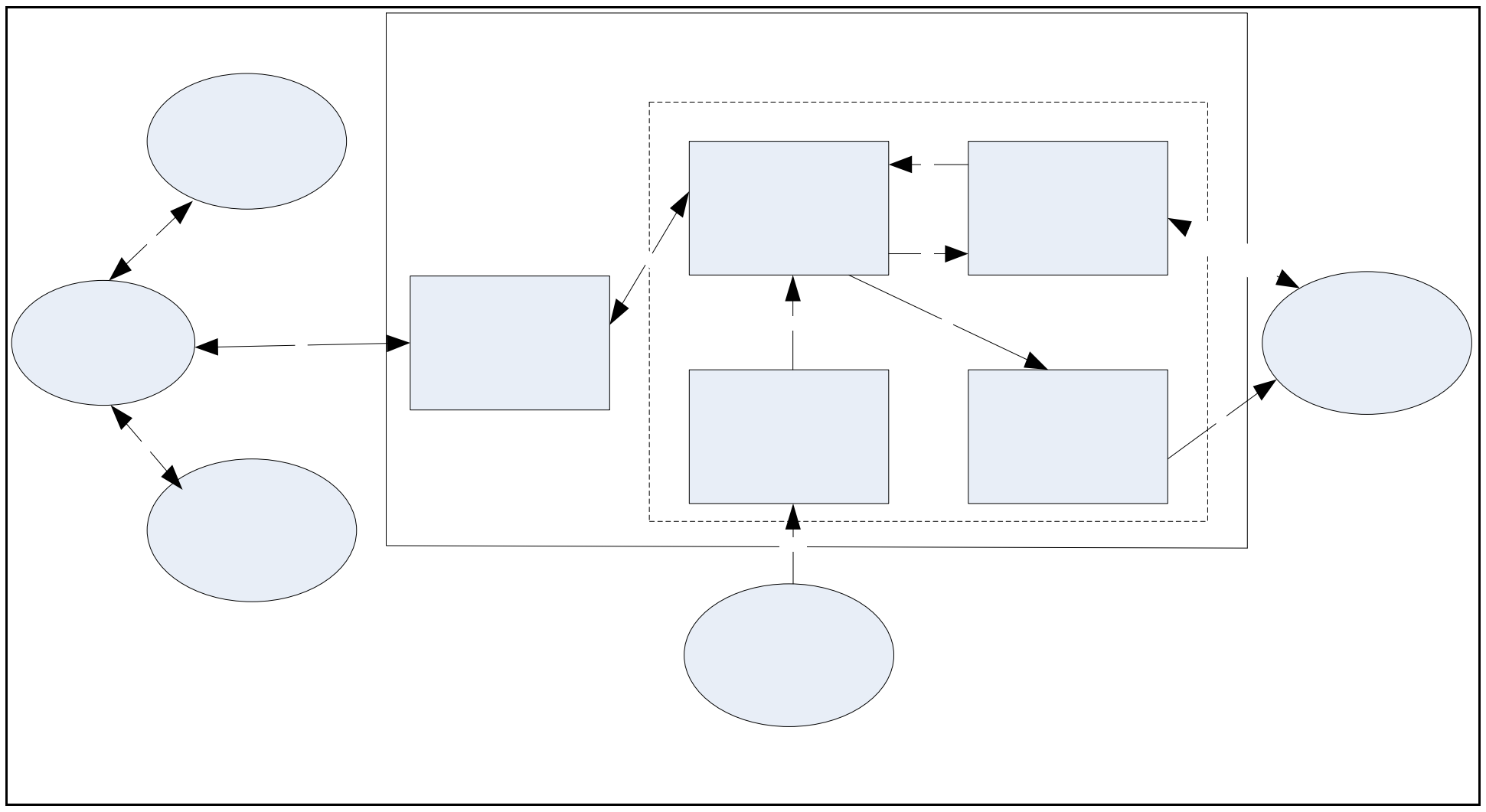

Figure 1. Proposed advisory system architecture

### *B. Cognition*

Problem solving varies in its external factors, including problem type and representation and internal characteristics of the problem solver. Structured and simple problems can be solved with regular rules and principles. They have knowable and comprehensible solutions where the relationship between decision choices and all problem states is known or probabilistic. Unstructured and complex problems possess multiple solutions, solution paths, or no solution at all. Unstructured problem possesses multiple criteria for evaluating solutions, so it is uncertain which concepts, rules, and principles are necessary for its solution and how they should be organized. It is often necessary for problem solvers to make judgments and express personal opinions or beliefs about the problem; so unstructured problems are uniquely human and interpersonal activities. Therefore, the frame or scenario-based case representation is suitable for well structured problem solving since the rules and principles of problem solving are well-defined. This means that the similar cases retrieved based on certain inputs or states can be applied to new problems. One of the knowledge acquisition frames designed for appointment of the demonstrator in university is shown in figure 2 .

```
        Knowledge acquisition frame 1
Appointment Of The Demonstrator
Certificates:(Bachelor):yes/No     Equivalent
yes/No
University (Recognized): yes/No
Estimation: good or higher
Study period: 4 or 5 or 6 or 7
Other conditions: age …
   The health situation …
   Marital status
Conditions of the Council of the dept.
Conditions of the Council of the faculty:
…………………………………………………………………
Conditions and exceptions of the Council of
the university:
……some…medical…specializations…………
Steady Committee for the appointment of
repeaters, lecturers, language teachers,
researchers assistants recommendation:
Yes/no
The opinion and recommendation of the
University Council:
Appoint the person
  Domain expert Name :
  Signature (     )
```

Figure 2. Frame for problem solving in the appointment of demonstrator

Different knowledge acquisition frames are designed to acquire knowledge in the different regulations of the university. The next stage is the knowledge representation.

### C. Representation of knowledge

One important class of architectural properties revolves around the representation of knowledge. Semantic networks, encodes both generic and specific knowledge in a declarative format that consists of nodes ( for concepts or entities) and links (for relations between them). Figure 3 shows the semantic network for the acceptance of new student in the university. Frames and schemas offer structured declarative formats to specify concepts in terms of attributes (slots) and their values (fillers).

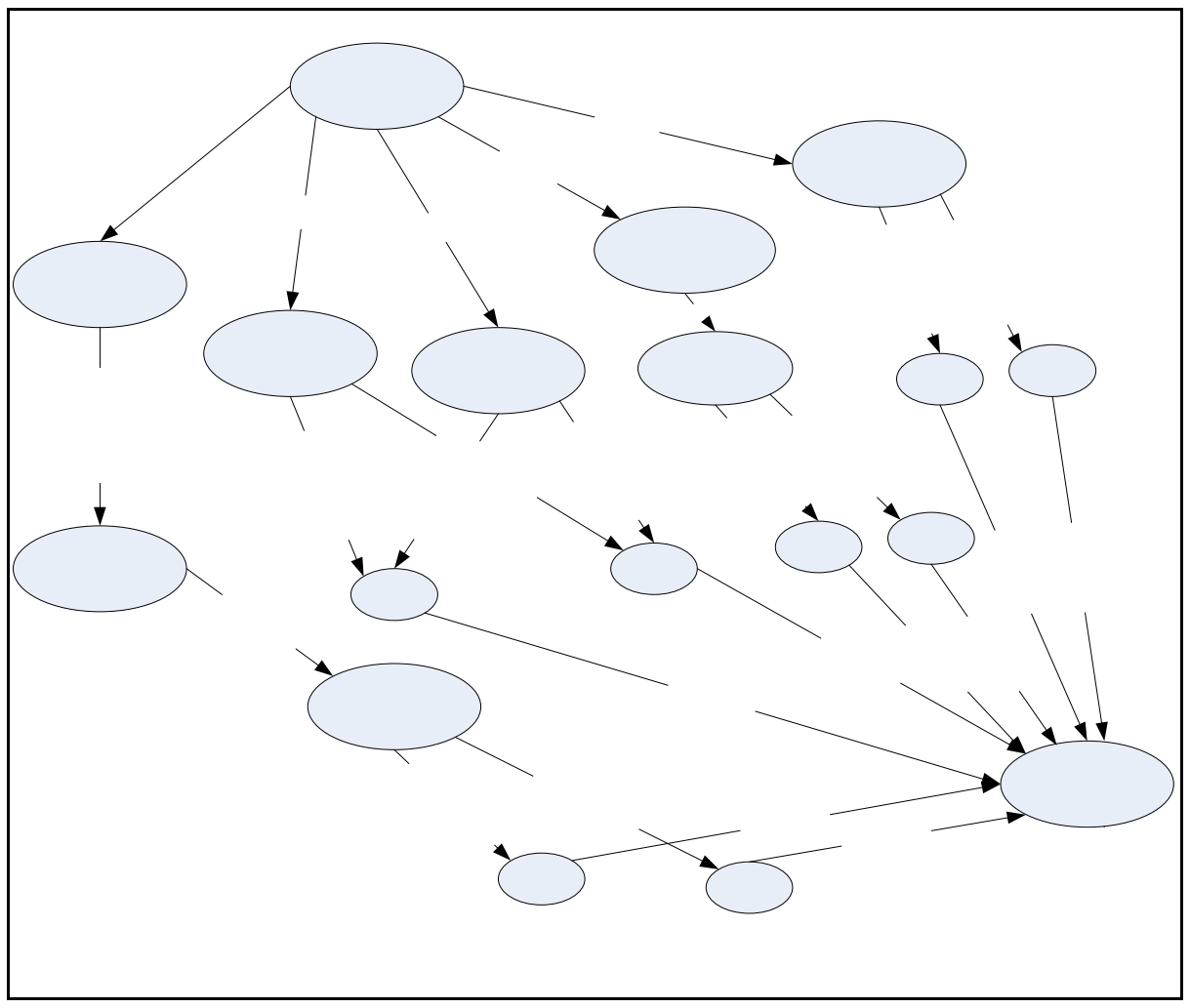

Figure 3. Semantic network for the acceptance of students in university

Table 1 shows the frames representing the semantic network shown in figure 3.

TABLE I. THE FRAMES OF STUDENTS IN UNIVERSITY

| Frame name | Slot | Slot value |
|---|---|---|
| Student | Has A | Behavior |
| | Has A | Certificate (education) |
| | Has A | Job |
| | Get | Personal interview |
| | Get | Health status |
| Behavior | Decision is | Not or OK |
| Certificate | Is | Up to date |
| Personal interview | Decision is | Not or OK |
| Health status | Decision is | Not or OK |
| Job | Belongs to | Affiliation |
| Affiliation | Approve | The study in university |
| The study in university | Decision is | Not or OK |
| OK | Give the | Legal authority |
| Not | Give the | Legal authority |
| Legal authority | - | - |

The flowchart shown in figure 4 explains the decisions applied for the acceptance of new student in university according the rules in the study and testing regulation.

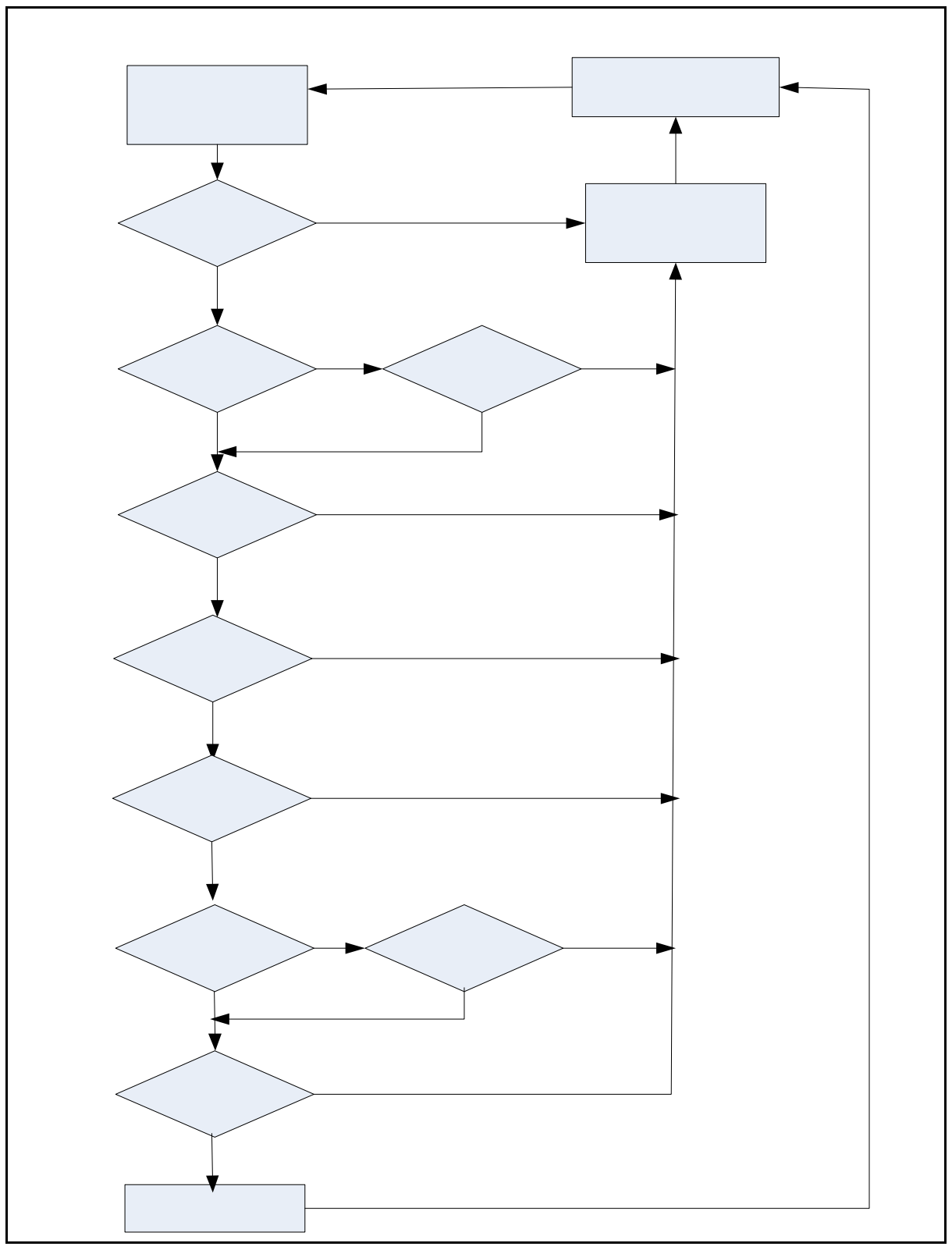

Figure 4. flowchart for accepting student in university

The knowledge base is implemented using CLIPS [16] . Sample of the rules included in the knowledge base are given in figure 5.

```
 (defmodule MAIN (export ?ALL))
     (defglobal ?*Decision_OK* = 0)
     ;0=No selection , 1=True selection,
              2=False selection
(defglobal ?*Decision_Causes* = "")
(defglobal ?*Decision_Law_Text* = "")
(defglobal ?*Decision_Law_Link* = "")
; definition of CLASSES
 (defclass MAIN::Final_Decision (is-a USER)
   (role concrete)
   (pattern-match reactive)
    (slot Decision_OK create-accessor read-
   write)(type INTEGER))  ;0=No selection,
               1=True, 2=False
   (slot   Decision_Causes   (create-accessor
read-write) (type STRING))
   (slot  Decision_Law_Text  (create-accessor
read-write) (type STRING))
  (slot   Decision_Law_Link   (create-accessor
read-write) (type STRING)))
;===== General Rules ===========
(defrule MAIN::List_Focus_01
   (List 01 ?n)
=>
   (switch ?n
       (case 01 then (focus LIST_01_01))
       (case 02 then (focus LIST_01_02))
       (case 03 then (focus LIST_01_03))
       (case 04 then (focus LIST_01_04))
       (case 05 then (focus LIST_01_05))
       (case 06 then (focus LIST_01_06))))
;====================
 (defrule MAIN::ConverFacts
 (SelGUI ?idx ?val ?ena ?stl ?tag)
=>
 (assert (Sel ?idx ?val ?ena ?stl ?tag)))
 (defmodule LIST_01_01 (import MAIN ?ALL)
 (export ?ALL))
;====================
  (defrule LIST_01_01::00 (declare (salience
                    100))
   (Sel ? ?val ?ena ?stl ?tag)
=>
    ;case of student acceptance
  (bind ?*Decision_Causes*"accept student")
  (bind      ?*Decision_Causes*      (str-cat
?*Decision_Causes*    "  The  differentiation
between  applicants,  who  apply  to  them  all
the conditions and according to their grades
in    the    secondary    school    certificate
test,personal  interview  and  admission  tests
if any. "))
(bind ?*Decision_Law_Text* "|rule3| rule 4")
(bind ?*Decision_Law_Link* "102-1-3|102-1-4"))
;====================
(defrule LIST_01_01::99(declare (salience -90))
   (Sel ? ?val ?ena ?stl ?tag)
=>
(make-instance CaseDecision of Final_Decision
       (Decision_OK ?*Decision_OK*)
       (Decision_Causes ?*Decision_Causes*)
       (Decision_Law_Text
?*Decision_Law_Text*)
       (Decision_Law_Link
?*Decision_Law_Link*)))
```

Figure 5. Samples of rules in the Knowledge base

## III. CASE STUDY

The higher education and universities council's law and its executives regulations in Saudi Arabia is a multi-criteria systems. It consists of 8 regulations. Each of them includes more than 7 subsystems. The number of rules in the regulations are listed in table 2.

TABLE II. RULES AND REGULATIONS OF HIGHER EDUCATION AND UNIVERSITIES COUNCIL'S LAW

| No | Regulation Name | Number of rules in regulation |
|---|---|---|
| 1 | Study and testing | 53 |
| 2 | Financial Affairs | 52 |
| 3 | The employment of non Saudis in the universities | 60 |
| 4 | Scholarships and training for the associates of universities | 41 |
| 5 | affairs of graduate study | 68 |
| 6 | Saudi university employees | 106 |
| 7 | Scientific Research | 51 |
| 8 | Scientific societies | 51 |

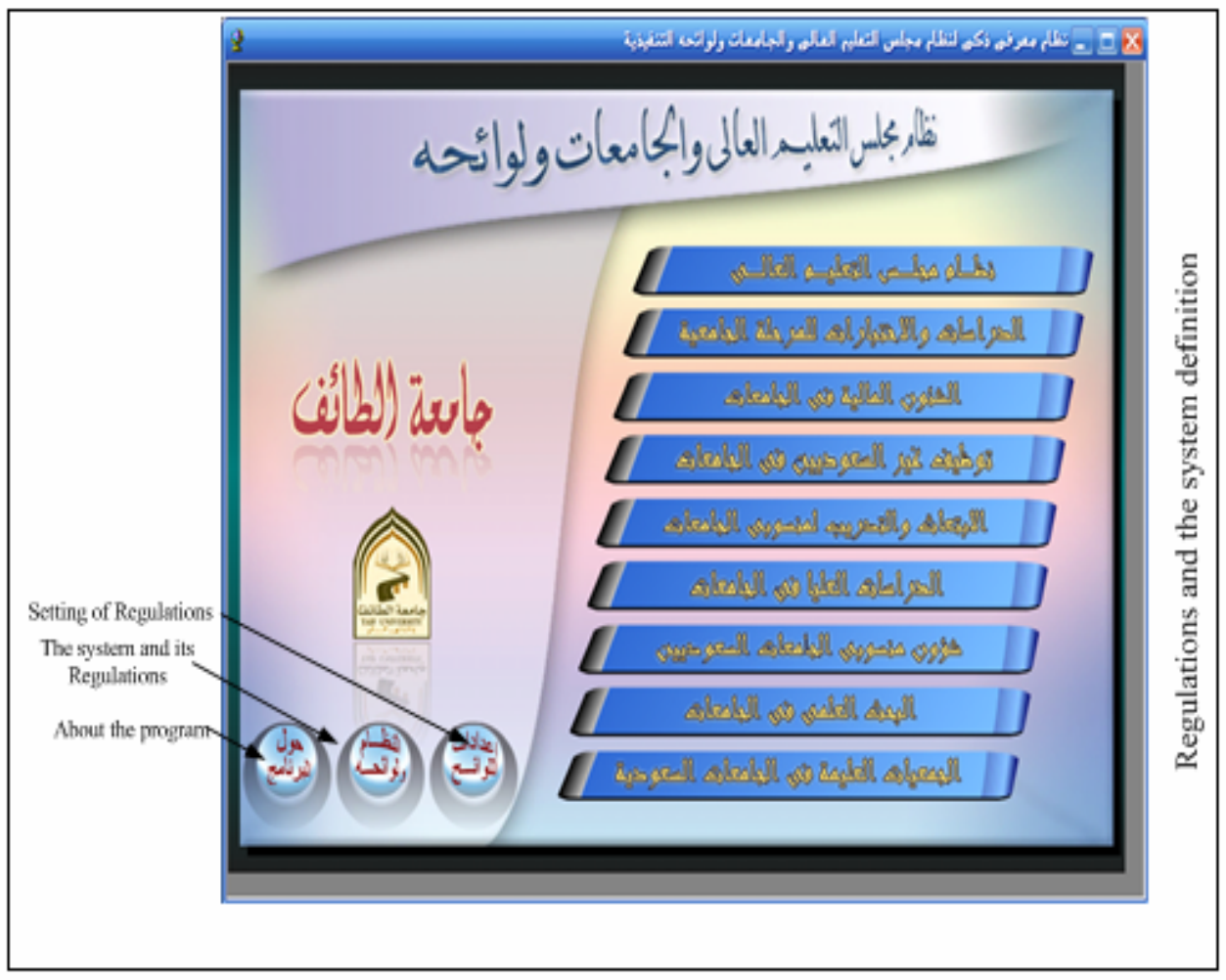

Figure 6. The main window of the proposed advisory system

The user interface of the proposed system is shown in figure 6. The decision making process in any subsidiary regulation needs series of queries. The answer to each query has a binary value yes or no. The answer in each case is followed by a decision or another query. All of these answers should be displayed in a main window and sometimes in accompanied dialogue window (exceptions). A part of this system is shown figure 7. The figure shows the decision and what are the rules that yield to the decision. Many other windows are developed for each criteria in the project.

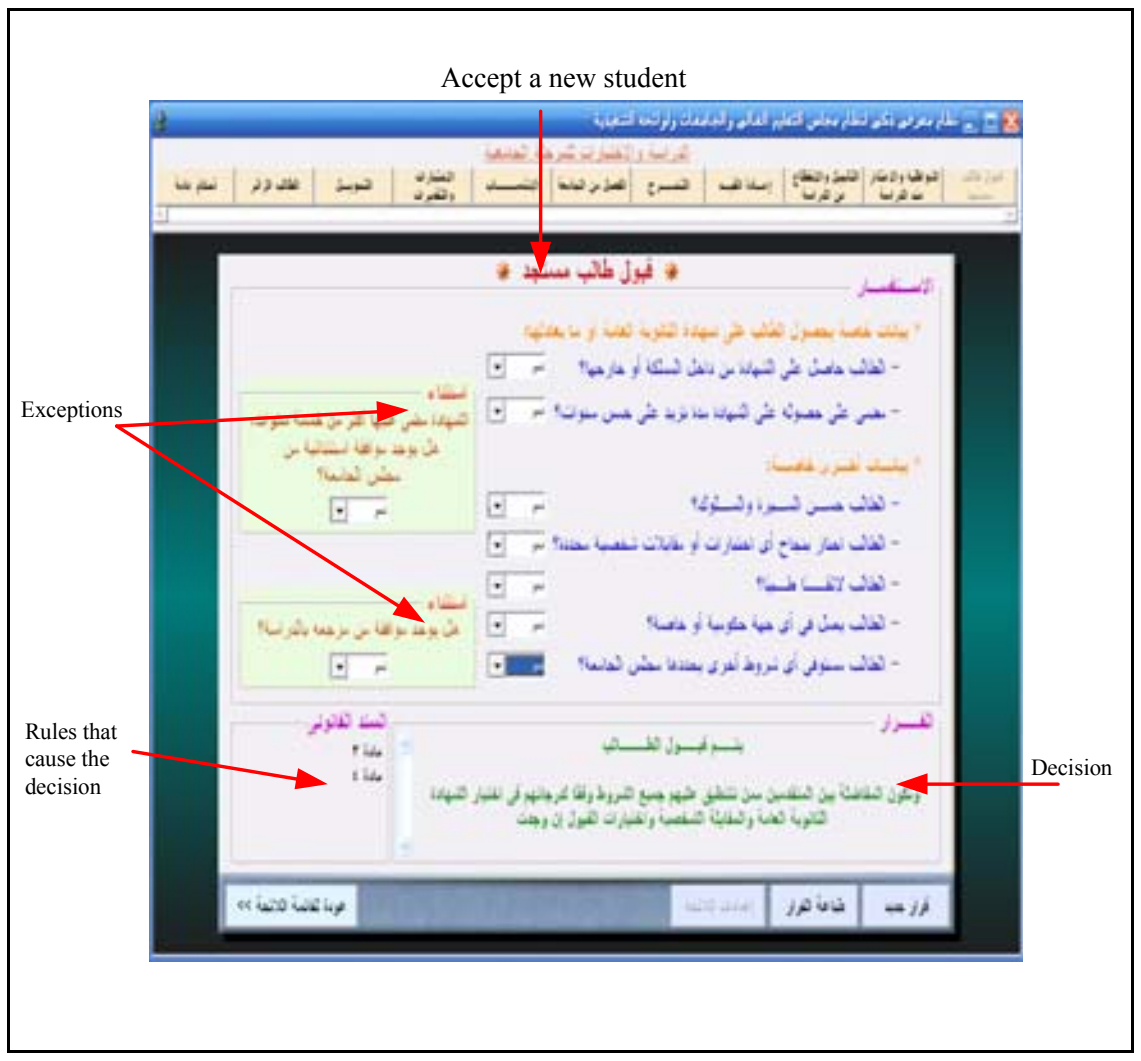

Figure 7. The decision , exception and the arguments window

## IV. SYSTEM FLEXIBILITY AND MERITS

Flexibility has become a key characteristic desired in both software systems and business processes. Software system flexibility is a two-dimensional construct composed of structural and process flexibility. Structural flexibility is the capability of the design and organization of a software application to be successfully adapted to business changes. Process flexibility is the ability of people to make changes to the technology using management processes that support business changes. The determinants of structural and process flexibility are based on measures of flexibility in the behavioral psychology and software engineering literature [17]. Change acceptance, modularity, and consistency are the measures used for structured flexibility in the proposed system. Change acceptance is the degree to which a system contains built-in capacity for change. Modularity is the degree of formal design separation within a software. Consistency is the degree to which data and components are integrated consistently across a software. The proposed system includes the possibility of amending some of the data that may occur in future, which assures the change acceptance. The system includes three main modules; scholarships and training, employment of non- Saudis and studies and tests, which assures the system modularity. Figure 8 shows both the change acceptance and the system modularity. The system consistency is assured by integrating the entire regulation and system definition of the Higher Education and Universities Council's Law and its Executives Regulations in Saudi Arabia as shown in figure 6. The process flexibility is measured by rate of response, expertise, and coordination of action. The proposed system accepts the changes that can be made in a timely manner that satisfies high rate of response. One of the major advantages of the proposed advisory system is its ability to up-to-date knowledge which yields to satisfy the expertise.

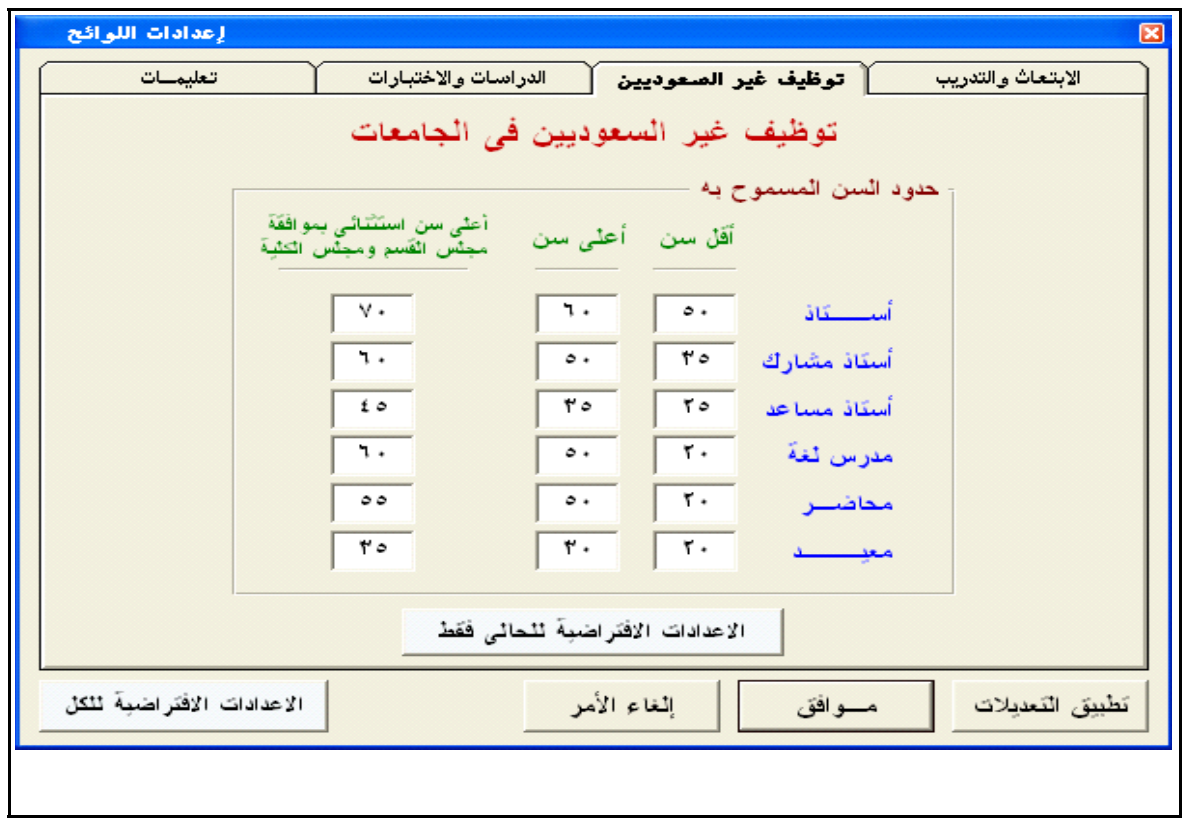

Figure 8. The seting window for the regualtion

## V. CONCLUSION AND FUTURE WORK

Intelligent advisory systems support decision maker in different domains specially in law. This paper presents an intelligent advisory system based on the executive regulations and rules that govern universities and institutes. This system provides legal advices to managers in universities and institutes. It does not substitute human advisors in law but it alleviates the burden based upon them. The advices are given automatically with the law causes and arguments. The system includes database which consists of a large number of rules and regulations. Also it is flexible enough to accept new setting without effecting the knowledge base.
Our future work will be concentrated on adopting the system to work online with different languages. Also, we will add additional knowledge in other domains to assist university managers.

## REFERENCES

[1] Leila Amgoud and Henri Prade "Using arguments for making and explaining decisions" Artificial Intelligence,vol. 173, pp. 413-436, 2009.

[2] Forslund, G., "Toward Cooperative Advice-Giving Systems: A Case Study in Knowledge Based Decision Support," *IEEE Expert*, pp. 56□-62,1995.

[3] Vanguard Software Corporation, Decision Script, 2006. Accessed via , www.vanguardsw.com/decisionscript/jgeneral.htm

[4] Aronson, J. and E. Turban, *Decision Support Systems and Intelligent Systems*. Upper Saddle River, NJ: Prentice-Hall, 2001.

[5] Magni, C.A., S. Malagoli and G. Mastroleo, "An Alternative Approach to Firms' Evaluation: Expert Systems and Fuzzy Logic," *Int J Int Tech Decis*, 5(1), 2006.

[6] Demir, C., S.H. Gultekin and B. Yener, "Learning the Topological Properties of Brain Tumors," *IEEE ACM T Comput*, 1(3), 2005.

[7] Gefen, D., E. Karahanna and D.W. Straub, “Trust and TAM in Online Shopping: An Integrated Model,” *MIS Quart*, 27(1), 2003.

[8] Luger G., *Artificial Intelligence: Structures and Strategies for Complex Problem Solving*. Addison Wesley, 2005.

[9] Gregg, D. and S. Walczak, “Auction Advisor: Online Auction Recommendation and Bidding Decision Support System,” *Decis Support Syst*, 41(2), 2006.

[10] Mintzberg, H., D. Raisinghani and A. Theoret, “The Structure of ‘Unstructured’ Decision Processes,” *Admin Sci Quart*, 21(2), 1976.

[11] Chandler, P. R. and M. Pachter, “Research Issues in Autonomous Control of Tactical UAVs,” in *Proceedings of the American Control Conference*, 1998.

[12] Sniezek J.A. and T. Buckley, “Cueing and Cognitive Conflict in Judge-Advisor Decision Making,” *Organ Behav Hum Dec*, 62(2), 1995.

[13] Arendt, L.A., R.L. Priem and H.A. Ndofor, “A CEO-Advisor Model of Strategic Decision Making,” *J Manage*, 31(5), 2005.

[14] Sniezek, J.A., “Judge Advisor Systems Theory and Research and Applications to Collaborative Systems and Technology,” in *Proceedings of the 32nd Hawaii International Conference on System Sciences*, 1999.

[15] http://www.utsystem.edu/News/mission.htm "University of Texas System.

[16] Joseph C. Giarratano "CLIPS User's Guide" version 6.2 March 31st 2002

[17] Nelson, K. M., and Ghods, M. .Evaluating the Contributions of a Structured Software Development and Maintenance Methodology,. Information Technology and Management, Winter 1998.